\documentclass[conference]{IEEEtran}
\IEEEoverridecommandlockouts
\usepackage{amsmath,amssymb,amsfonts}

\usepackage{graphicx}
\usepackage{textcomp}
\usepackage{xcolor}
\usepackage{subfig}
\usepackage{textcomp}
\usepackage{xcolor}
\usepackage{tikz}
\usepackage{quantikz}
\usepackage{algorithm}
\usepackage{algorithmicx}
\usepackage{float}

\usepackage{algpseudocode}%

\usepackage{pgffor}
\usepackage{soul}
\usepackage[sorting=none, doi = false, isbn = false, style = ieee]{biblatex} % for a biblio
\usepackage{csquotes}
\usepackage{hyperref}% add hypertext capabilities
\usepackage{float}
\usepackage[english]{babel}
\usepackage{amsthm}
\usepackage{amssymb}
\usepackage{enumitem}
\usepackage{dsfont}
\usepackage{tikz}
\usepackage{multirow}%
\usepackage{subfig}
\usepackage{bm}
\usepackage{bbold}

\usepackage[a4paper, total={7in, 10in}]{geometry}

\usetikzlibrary{shapes.geometric, arrows, quantikz2}

\def\BibTeX{{\rm B\kern-.05em{\sc i\kern-.025em b}\kern-.08em
    T\kern-.1667em\lower.7ex\hbox{E}\kern-.125emX}}

\begin{document}

\title{Quantum Annealing for Active User Detection in NOMA Systems}

\author{\IEEEauthorblockN{Romain Piron}
\IEEEauthorblockA{\textit{Universit\'e de Lyon, INSA Lyon, INRIA} \\
\textit{CITI EA 3720}\\
69621 Villeurbanne, France \\
romain.piron@insa-lyon.fr}
\and
\IEEEauthorblockN{Claire Goursaud}
\IEEEauthorblockA{\textit{Universit\'e de Lyon, INSA Lyon, INRIA} \\
\textit{CITI EA 3720}\\
69621 Villeurbanne, France \\
claire.goursaud@insa-lyon.fr}

}

\maketitle

\begin{abstract}
Detecting active users in a non-orthogonal multiple access (NOMA) network poses a significant challenge for 5G/6G applications. Traditional algorithms tackling this task, relying on classical processors, have to make a compromise between performance and complexity. However, a quantum computing based strategy called quantum annealing (QA) can mitigate this trade-off. In this paper, we first propose a mapping between the AUD searching problem and the identification of the ground state of an Ising Hamiltonian. Then, we compare the execution times of our QA approach for several code domain multiple access (CDMA) scenarios. We evaluate the impact of the cross-correlation properties of the chosen codes in a NOMA network for detecting the active user's set.

\end{abstract}


\section{Introduction}

In 6G wireless networks, particularly for Industrial Internet of Things, a large number of user equipment (UE) aim at transmitting data to an access point (AP) with strict reliability and latency requirements \cite{chetot_active_2023}. Instead of assigning time-slot frequencies to each node, resources are allocated only to active UEs in such massive networks \cite{chetot_active_2023, jeannerot_mitigating_2023}. Consequently, identifying these active UEs at the start of each frame is crucial and is known as active user detection (AUD).

To do so, each active UE sends a pilot sequence to the AP to indicate its activity. Leveraging non-orthogonal multiple access (NOMA) recently appeared relevant in this context \cite{dai_survey_2018} but it introduces interference between the pilots. The most reliable, but also the most costly, way to perform AUD in such wireless networks is a maximum-likelihood (ML) estimation of the activity pattern \cite{chetot_activity_nodate}.

Two main approaches allow to formulate this ML estimation \cite{wang_covariance-based_2023}. The first one consists in jointly estimating the channel and the activity pattern \cite{chetot_active_2023} which requires to compute simultaneously both estimators of the channel matrix and of the active UE. The second approach is commonly referred as \textit{covariance based approach} \cite{wang_covariance-based_2023, marata_activity_2024} and only requires to know the statistics of the channel matrix but not its realization. 

Inspired by the covariance based approach, a recent technique proposed in \cite{haghighatshoar_improved_2018} formulates AUD as a least square problem. Such formulation nicely falls into the general class of \textit{quadratic binary unconstrained optimization} (QUBO) problems by assuming unit large-scale fadings. QUBO problems are known to be NP-hard \cite{haghighatshoar_improved_2018} since an instance of size $N$ requires $\mathcal{O}\left(2^N\right)$ operations to be solved. 

Nevertheless, the emergence of quantum algorithms \cite{boixo_quantum_2014} offers an interesting perspective to deal with this issue. Indeed, quantum annealing (QA) \cite{kadowaki_quantum_1998, farhi_quantum_2000} is a quantum computing-based strategy promising for solving QUBO problems with a reduced execution time. It has gained significant interest over the past two decades and already outperformed classical thermal annealing for some instances of QUBO problems \cite{hauke_perspectives_2020}. It is why we proposed in \cite{piron_scheduling_2024} an adaptation of QA for AUD in a simple scenario where the AP knows the channel realizations. In this work, we aim to to go further by considering a more realistic scenario. Our contributions are as follows:
\begin{itemize}
    \item We show how to build an Ising Hamiltonian associated to the covariance based AUD problem. To do so, we use the least square approach introduced in \cite{haghighatshoar_improved_2018}. This Hamiltonian allows to parameterize a QA approach for computing an estimator of the activity pattern of the network
    \item The reliability of our QA-detector is measured with the \textit{quantum activity error-rate} (qAER) that we introduce
    \item We show that the choice of the random scheme to design the pilot matrix has an impact on the time taken by the QA approach to converge towards an acceptable qAER
\end{itemize}
The rest of the paper is organized as follows. Sec. \ref{problem_def} defines the signal model and presents the least square approach we use. Sec \ref{construct_QA} presents quantum annealing and the construction of our Ising Hamiltonian. Our metrics and our numerical results are presented in Sec. \ref{analysis_algo}.

\subsection*{Notations}
\hspace{-0.5cm}
\begin{tabular}{|ll|}
    \hline
    \textit{Object} \hspace{5cm} & \textit{Notation} \\
    \hline
     Scalar &  $x$  \\
     Vector & $\bm{x}$ \\
     Matrix & $\bm{X}$ \\
     Diagonal matrix such that $X_{ii} = x_i$ & $\text{Diag}(\bm{x})$ \\
     Sphere of radius 1 in $d$ dimensions & $S^{d-1}$ \\
     $l_2$ norm & $\left\lVert . \right\rVert_2$ \\
     Froebenius norm &  $\left\lVert . \right\rVert_F$ \\
     Determinant of a matrix & $|\bm{X}|$ \\
     Expectated value with respect to the density of $x$ & $\mathbb{E}_x (\cdot)$ \\
     Gaussian law of mean $\bm{\mu}$ and covariance matrix $\bm{\Sigma}$ & $\mathcal{N}\left(\bm{\mu}, \bm{\Sigma}\right)$ \\
     \hline
\end{tabular}

\section{Problem definition}
\label{problem_def}
\subsection{System model}

We consider the non-coherent SIMO model of \cite{ngo_multi-user_2020}. In this scenario, $N$ users are equipped with a single antenna and assigned with a specific pilot sequence $\bm{p}_i \in \mathbb{C}^M$ with $M < N$. We assume that the power transmitted by each user is normalized on average. 
\begin{align}
    \mathbb{E}_{\bm{p}_i}\left(\left\lVert \bm{p}_i \right\rVert ^2 \right) = 1 \,\, \text{for} \,\, i = 1,\dots,N
\end{align}

At the beginning of the frame, a random subset of these users transmit their sequence to a $K$-antenna access point (AP). It allows to define the \textit{activity pattern} of the network $\bm{\alpha}^{(0)} \in \{0,1\}^N$ as:
\begin{align}
    \alpha^{(0)}_i = \begin{dcases}
        1 & \, \text{if } i\text{-th user is active} \\
        0 & \, \text{otherwise}
    \end{dcases}, \,\, i = 1, \dots, N
\end{align}
The channel vector $\bm{h}_i \in \mathbb{C}^K$ between the $i$-th user and the AP is modelled by a Rayleigh fading law with a unitary large scale fading coefficient which means:
\begin{align}
    \bm{h}_i \sim \mathcal{N}(0, I_K) \,\,\, i = 1, \dots, N
\end{align}
Furthermore, each antenna suffers from an additive white Gaussian noise $\bm{w}_k \in \mathbb{C}^M$(AWGN):
\begin{align}
    \bm{w}_k \sim \mathcal{N}(0,\xi^2 I_M)
\end{align}
Thus, the received signal at the antenna reads:
\begin{align}
    \begin{split}
    \bm{Y} &= \sum_{i=1}^N \alpha_i^{(0)} \bm{p}_i \bm{h}_i^T + \bm{W} \\
    &= \bm{P}.\text{Diag}\left(\bm{\alpha}^{(0)}\right). \bm{H}^T + \bm{W},
    \end{split}
\end{align}
where we introduced:
\begin{align}
\begin{split}
    \bm{P} = [\bm{p}_1 \cdots \bm{p}_N] \in \mathbb{C}^{M\times N} & \text{ (Pilot matrix)} \\
    \bm{H} = [\bm{h}_1 \cdots \bm{h}_N] \in \mathbb{C}^{K \times N} & \text{ (Channel matrix)} \\
    \bm{W} = [\bm{w}_1 \cdots \bm{w}_K] \in \mathbb{C}^{M \times K} & \text{ (AWGN)}
\end{split}
\end{align}
Under our assumptions, the signal-to-noise ratio (SNR) of the transmitted signal is the same for every user \cite{ngo_multi-user_2020} and is expressed in dB as:
\begin{align}
    \text{SNR} = -10\log_{10} \left(M\xi^2\right)
    \label{snr_def}
\end{align}
In this work, we limit our study to small system sizes. Thus, we fix once for all $N = 5$ and $M = 4$.

\subsection{Covariance-based activity detection}

Let us now briefly review the main ingredients beyond the so-called covariance based approach for AUD. We assume that there is no correlation between the antennas of the AP. Then, it has been shown \cite{haghighatshoar_improved_2018, yu_massive_nodate} that, given a certain activity pattern $\bm{\alpha}$, each column of the received signal $\bm{y}_k$ follows the same Gaussian law $\mathcal{N}\left(\bm{0}, \bm{\Sigma}\right)$. An estimator of $\bm{\Sigma}$ is given by the sample covariance matrix \cite{haghighatshoar_improved_2018, wang_covariance-based_2023}:
\begin{align}
    \hat{\bm{\Sigma}}_Y = \frac{1}{K} \sum_{k=1}^K \bm{y}_k \bm{y}_k^H = \frac{1}{K} \bm{Y} \bm{Y}^H
\end{align}
The authors of \cite{haghighatshoar_improved_2018} show that $\hat{\bm{\Sigma}}_Y$ is a sufficient statistics for the activity pattern recovery. Thus, it motivates the introduction of their non-negative least square (NNLS) approach which consists in matching the true covariance matrix $\bm{\Sigma}$ in the absence of additive noise to the sample covariance matrix \cite{haghighatshoar_improved_2018}. Given our assumption on the large scale fading coefficients, the NNLS estimator nicely recasts in the following combinatorial optimization problem:
\begin{align}
    \bm{\alpha}^{\text{NNLS}} = \underset{\bm{\alpha} \in \{0,1\}N}{\text{arg min }} \left\lVert \bm{P}.\text{Diag}(\bm{\alpha}).\bm{P}^H - \hat{\bm{\Sigma}}_{Y} \right\rVert_F^2
    \label{NNLS_min_pb}
\end{align}
Such formulation has the advantage to not require the knowledge of the noise level $\xi^2$, which explains that NNLS has gained interest for signal processing applications \cite{haghighatshoar_improved_2018}.

Furthermore, Eq. \ref{NNLS_min_pb} is exactly a quadratic unconstrained binary optimization (QUBO) problem \cite{hauke_perspectives_2020} . Thus, another significant advantage of the NNLS detector is that it is immediately suited for algorithmic approaches known to solve QUBO problems. Thus, quantum annealing can be used for activity detection in this context. \cite{hauke_perspectives_2020}

\subsection{Pilot scheme}
The design of the pilot matrix $\bm{P}$ impacts the structure of the objective function involved in \ref{NNLS_min_pb}. In this work, we aim to evaluate whether the choice of this design has a strong impact on the efficiency of a QA algorithm for the activity pattern recovery.

Notice that we do not aim at studying the recovery performance of NNLS with respect to the design of $\bm{P}$ but rather the efficiency of our QA algorithm for computing $\bm{\alpha}^{\text{NNLS}}$. Hence we do not deeply delve into mathematical aspects concerning the guarantees of recovery of the NNLS estimator \cite{haghighatshoar_improved_2018}. 

For the sake of simplicity, we propose the two random design schemes:
\begin{align}
    \bm{p}_i \sim \begin{dcases}
        \mathcal{N}\left(0, \frac{1}{\sqrt{M}}I_M\right) & \text{(Gaussian)} \\
        \mathcal{U}\left(S^{M-1}\right) & \text{(Unit sphere)}
    \end{dcases}, \,\, i = 1, \dots, N
\end{align}
referred respectively as \textit{Gaussian scheme} and \textit{unit sphere scheme}.

\subsection{Asymptotic case of large number of antenna}

The asymptotic case of a huge number of antenna $K \gg 1$ is interesting for us. Indeed, \cite{haghighatshoar_improved_2018} underlines the property:
\begin{align}
    \hat{\bm{\Sigma}}_{Y} \underset{K \rightarrow \infty}{\longrightarrow} \bm{\Sigma}
\end{align}
Thus, one can legitimately expect a perfect recovery from the NNLS estimator in the limit $K \rightarrow \infty$. This behavior can be observed through the \textit{activity error-rate} \cite{chetot_activity_nodate} of the decoder:
\begin{align}
    \text{AER}^{\text{NNLS}} = \frac{1}{N} \mathbb{E}_{(\bm{P},\bm{Y})}\left(\sum_{i=1}^N \delta \left( \alpha_i^{\text{NNLS}} \ne \alpha_i^{(0)} \right) \right),
\end{align}
which quantifies the reliability of $\bm{\alpha}^{\text{NNLS}}$.

\begin{figure}
    \centering
    \includegraphics[width=0.7\linewidth]{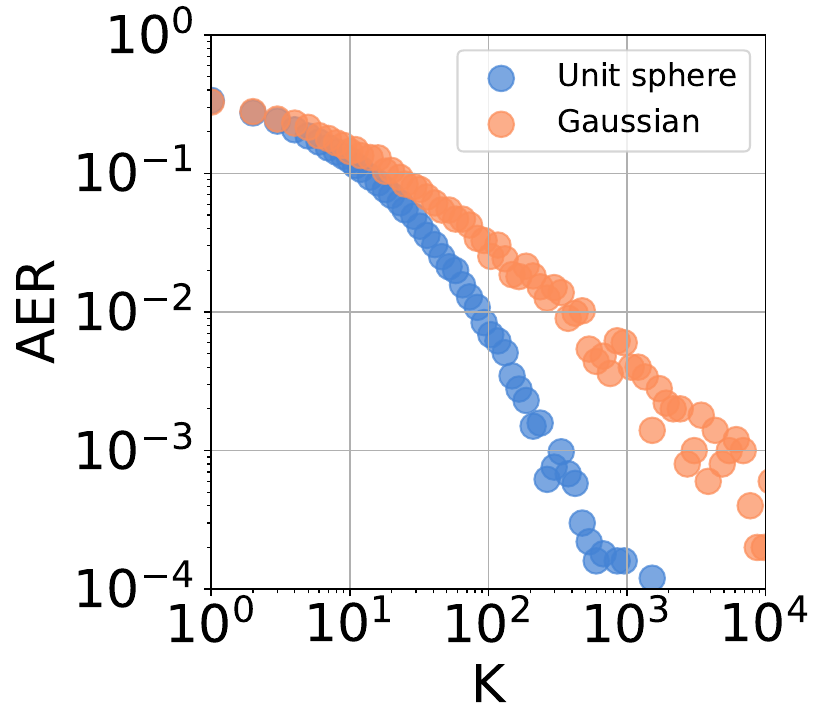}
    \caption{$\text{AER}^{\text{NNLS}}$ against the number of antenna $K$ at the AP for Gaussian and unit sphere schemes, at $\text{SNR} = \infty$}
    \label{fig:BER_comparison_N5M4}
\end{figure}

We reported on Fig. \ref{fig:BER_comparison_N5M4} the shape of $\text{AER}^{\text{NNLS}}$ against $K$ for our two coding schemes with $\xi^2 = 0$. The plot clearly shows that $\text{AER}^{\text{NNLS}} \rightarrow 0$ when $K \rightarrow \infty$. Since we do not focus on the intrinsic recovery properties of the NNLS detector, we adopt this asymptotic regime and choose $K$
so that $\text{AER}^{\text{NNLS}}\left(\text{SNR} \rightarrow \infty\right) \sim 10^{-3}$ for both schemes.

\section{Quantum annealing for NNLS computation}
\label{construct_QA}
Now that the wireless system is defined, we are ready to build a quantum annealing algorithm to compute the NNLS detector.

\begin{figure*}
    \centering
    \subfloat[Mean gaps \label{subFig:mean_gaps}]{%
    \includegraphics[scale=0.35]{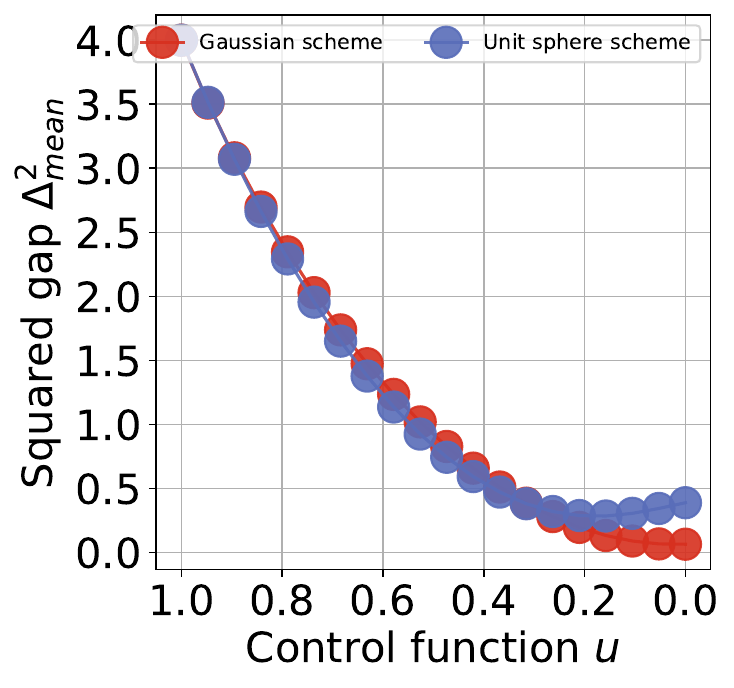}} \hspace{1cm}
    \subfloat[Gaussian scheme ($\epsilon = 0.01$) \label{subFig:overlap_control_example_gaussian}]{%
    \includegraphics[scale=0.4]{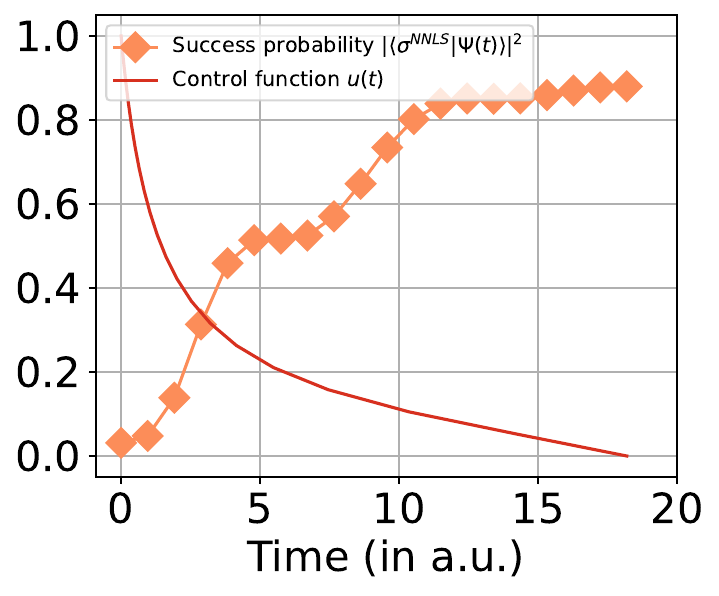}}\hspace{1cm}
    \subfloat[Unit sphere scheme ($\epsilon = 0.1$) \label{subFig:overlap_control_example_unit_sphere}]{%
    \includegraphics[scale=0.4]{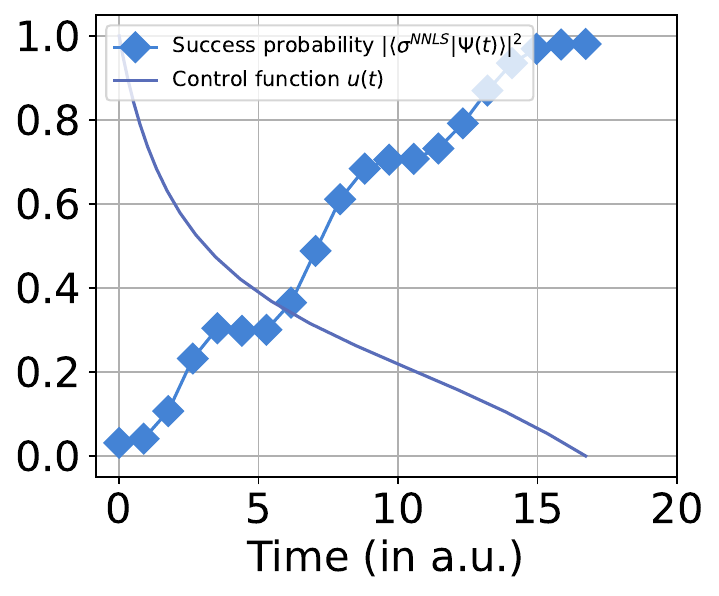}}
    \caption{The mean gaps for both pilot designs (a) allow to compute the mean control functions of Eq. \ref{ODE_u_mean} and evolve the state of the system towards $\ket{\bm{\sigma}^{\text{NNLS}}}$(b,c). The current state of the system is denoted $\ket{\Psi(t)}$, naturally the final state is $\ket{\Psi(T)}$.} 
  \label{fig:example_QA_approach}    
\end{figure*}

\subsection{Construction of the Ising Hamiltonian}

Quantum annealing is in fact well suited for the minimization of energy functions called \textit{Ising Hamiltonians} \cite{kadowaki_quantum_1998}. The first task to do is then to map the objective function of the minimization problem to an Ising Hamiltonian. Fortunately, such mapping has been studied when one deals with QUBO problems \cite{hauke_perspectives_2020}. To do so, we introduce the \textit{spin configuration} associated to an activity pattern $\bm{\alpha}$  \cite{piron_scheduling_2024}:
\begin{align}
    \bm{\sigma} = 1 - 2 \bm{\alpha} \in \{-1 , 1\}^N
\end{align}
Then, one can write the squared norm involved in the NNLS detector (Eq. \ref{NNLS_min_pb}) as a function of the spin variables $\sigma_i$. Throwing away the constant terms, we obtain following Ising Hamiltonian known as the \textit{problem Hamiltonian}:
\begin{align}
    H_P(\bm{\sigma}) = - \sum_{i < j } J_{ij} \sigma_i \sigma_j - \sum_i b_i \sigma_i
    \label{problem_hamilt}
\end{align}
where the coupling parameters $J_{ij}$ and the local fields $b_i$ are defined by :
\begin{align}
\begin{dcases}
    b_i &= - \text{Tr}\left(\bm{p}_i \bm{p}_i^H \hat{\bm{\Sigma}}_Y\right) + \frac{1}{2} \sum_{j=1}^N |\bm{p}_i^H \bm{p}_j|^2\\
    J_{ij} &= -\frac{1}{2} |\bm{p}_i^H \bm{p}_j|^2
    \end{dcases},
    \label{isingParameters}
\end{align}
The spin configuration that minimizes $H_P$ is called the \textit{ground state} (GS) of $H_P$. We denote it $\bm{\sigma}^{\text{NNLS}}$ as it exactly corresponds to the spin configuration associated to $\bm{\alpha}^{\text{NNLS}}$. Since we adopted the regime $K \gg 1$, we can reasonably consider that in most cases $\bm{\sigma}^{\text{NNLS}} = \bm{\sigma}^{(0)}$ as explained previously.

\subsection{Basics of QA}

Each node of the network is now assigned with a qubit labelled by the same index $i$. We promote each classical spin $\sigma_i$ to quantum spins described by the $z$ Pauli matrices $\hat{\sigma}_i^z$ \cite{piron_scheduling_2024} in order to encode the problem into the following quantum operator:
\begin{align}
    \hat{H}_P = - \sum_{i < j } J_{ij} \hat{\sigma}^z_i \hat{\sigma}^z_j - \sum_i b_i \hat{\sigma}^z_i.
\end{align}
The eigenstates $\ket{\bm{\sigma}}$ of $\hat{H}_P$ correspond to the classical configurations $\bm{\sigma}$ and the eigenvalues are given by the energy levels $H_P(\bm{\sigma})$ of the classical Hamiltonian. QA uses a transverse term $\hat{H}_C$ called the \textit{control Hamiltonian} in order to introduce quantum fluctuations between the states $\ket{\bm{\sigma}}$. These fluctuations allow to explore the different classical configurations in order to converge towards $\ket{\bm{\sigma}^{\text{NNLS}}}$.

We adopt the transverse control Hamiltonian used in \cite{kadowaki_quantum_1998}:
\begin{align}
    \hat{H}_C = - \sum_i \hat{\sigma}^x_i,
\end{align}
The QA algorithm relies on the global Hamiltonian:
\begin{align}
    \hat{H}(u(t)) = (1-u(t)) \hat{H}_P + u(t) \hat{H}_C,
\end{align}
initialized with $u(0) = 1$. The system is prepared in the ground state of $\hat{H}_C$ which is known. Then, one slowly decreases $u(t)$ so that the system remains close from its instantaneous GS. After a certain time $T$ called the \textit{annealing time}, one ends up with a final state denoted $\ket{\Psi(T)}$ that is expected to be close to $\ket{\bm{\sigma}^{\text{NNLS}}}$. The overlap between the two quantum states $|\langle \Psi(T) | \bm{\sigma}^{\text{NNLS}} \rangle |^2$ corresponds to the success probability of the algorithm. Let us now determine the time evolution of $u$.

\subsection{Scheduling with the mean gap}

The term $u$ is called the \textit{control function} and its time evolution is controlled by the squared gap:
\begin{align}
    \Delta_{(\bm{P},\bm{Y})}(u(t))^2 = \left(\varepsilon_1(u(t)) - \varepsilon_0(u(t))\right)^2 \,\,\,\, u(t) \in [0,1], 
    \label{spectral_gap}
\end{align}
where $\varepsilon_0$ is the lowest energy of the spectrum (lowest eigenvalue) of $\hat{H}$ and $\varepsilon_1$ is the first excited energy level. The $(\bm{P},\bm{Y})-$dependency comes from the expressions of the $J_{ij}'s$ and the $b_i$'s in Eq. \ref{isingParameters}. However, it would be inconvenient to require to compute this gap for each pair $(\bm{P},\bm{Y})$ which is why we introduce the \textit{mean gap} \cite{piron_scheduling_2024} as:
\begin{align}
        \Delta_{(M,N)}^2 &= \mathbb{E}_{(\bm{P},\bm{Y})} \left(\Delta^2_{(\bm{P},\bm{Y})} | \bm{W} = 0 \right).
\end{align}
This mean gap only depends on the pilot design and allows to compute the control function by solving the differential equation \cite{piron_scheduling_2024}:
\begin{align}
    \begin{dcases}
        \frac{du}{dt} &= -\epsilon \Delta^2_{\text{mean},(M,N)}(u(t)) \\
        u(0) &= 1
    \end{dcases},
    \label{ODE_u_mean}
\end{align}
where $\epsilon > 0$ is an arbitrary small constant called the precision level. The annealing time taken by the control function to reach its final value $u(T) = 0$ is obtained by integrating both sides of the previous equation:
\begin{align}
    T = \frac{1}{\epsilon} \int_0^1 \frac{du}{\Delta^2_{\text{mean},(M,N)}(u)}
\end{align}
Clearly, decreasing the precision level $\epsilon$ increases the annealing time. However, it also ensures $|du/dt| \ll \Delta^2$ which pushes the success probability close from 1 as explained in \cite{piron_scheduling_2024}. 

Given a problem instance $(\bm{P}, \bm{Y})$, one can compute the gap of the Hamiltonian $\hat{H}(u(t))$ at specific points $u \in [0,1]$ without even knowing the time dependency $u(t)$. We did so for several problem instances in order to estimate the mean gaps associated to both Gaussian scheme and unit sphere scheme. We reported on Fig. \ref{subFig:mean_gaps} the obtained shapes against $u$ which is used an affine parameter. The two schemes exhibit different behaviors of their associated mean gaps, which will yield to different control functions.

In order to compute them, we choose a precision level $\epsilon$ and solve Eq. \ref{ODE_u_mean}. For our two schemes, we executed a QA algorithm parameterized with the appropriated control function for the NNLS recovery associated to an instance $(\bm{P},\bm{Y})$. Fig. \ref{subFig:overlap_control_example_gaussian} and \ref{subFig:overlap_control_example_unit_sphere} show the mean control function and the evolution of the success probability for each evolution. The success probabilities are close from 1 at the end of the process as expected. 

\section{Analysis of the algorithm}
\label{analysis_algo}
The parameterization we introduced relies on the mean gap taken with $\bm{W} = 0$. We should now check the robustness of the approach against the SNR. Then, we will finally be able to compare the two schemes in terms of execution time. 

\begin{figure*}
\centering
    \subfloat[Gaussian  scheme \label{subFig:qBER_vs_SNR_gaussian_N5M4}]{%
    \includegraphics[scale=0.28]{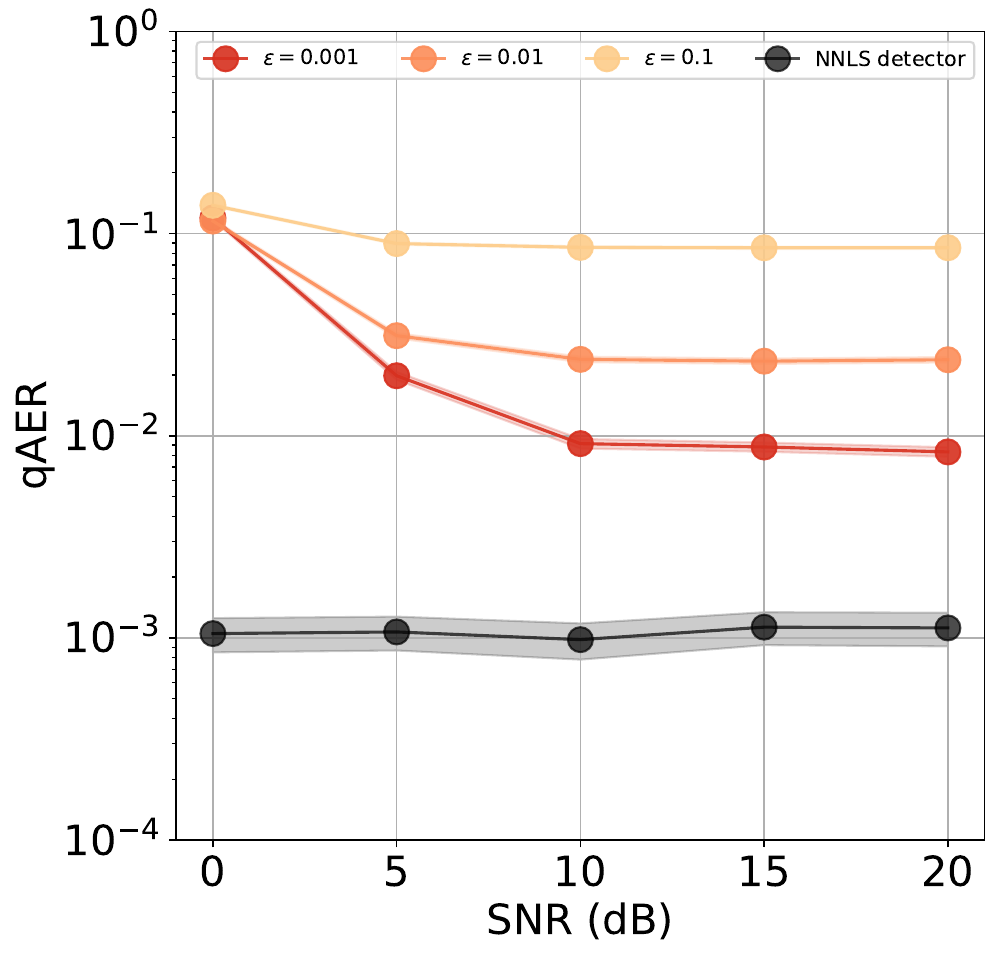}} \hspace{1cm}
    \subfloat[Unit sphere scheme \label{subFig:qBER_vs_SNR_unit_sphere_N5M4}]{%
    \includegraphics[scale=0.28]{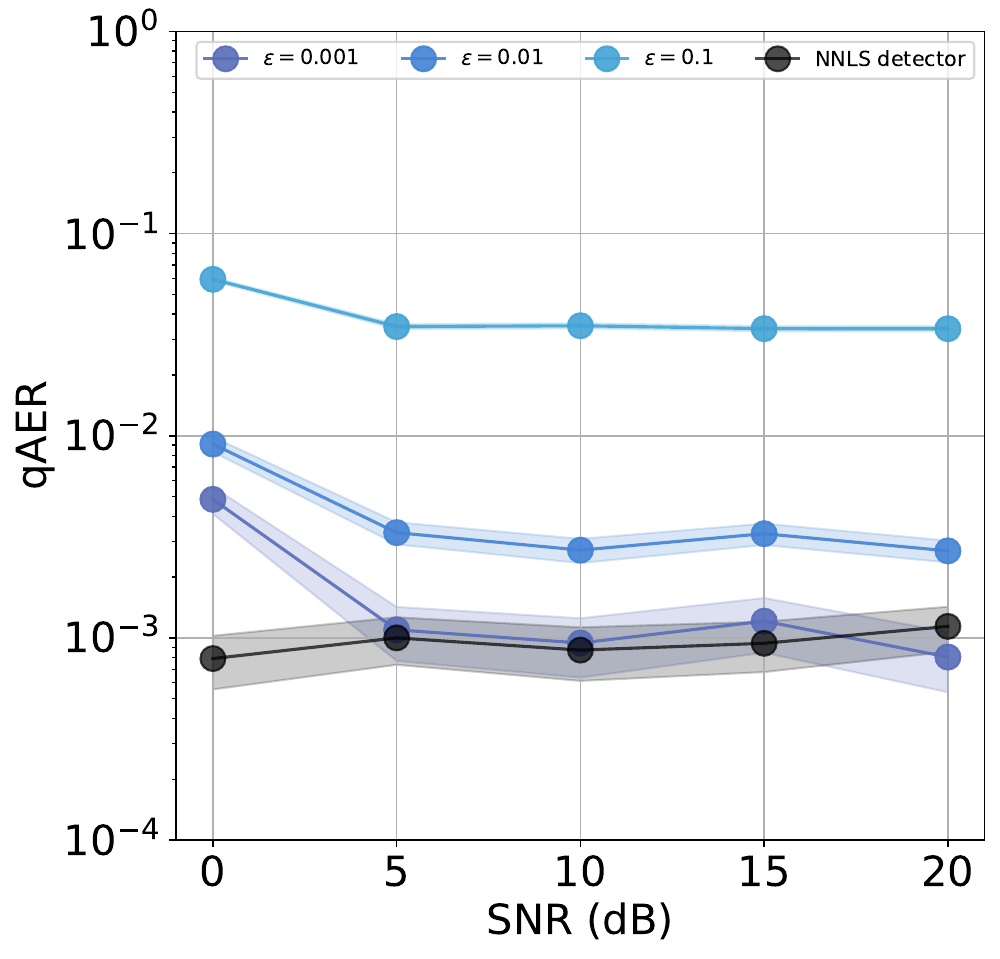}} \hspace{1cm}
    \subfloat[Annealing time analysis \label{subFig:qBER_vs_annealing_time_N5M4_snr10}]{%
    \includegraphics[scale=0.28]{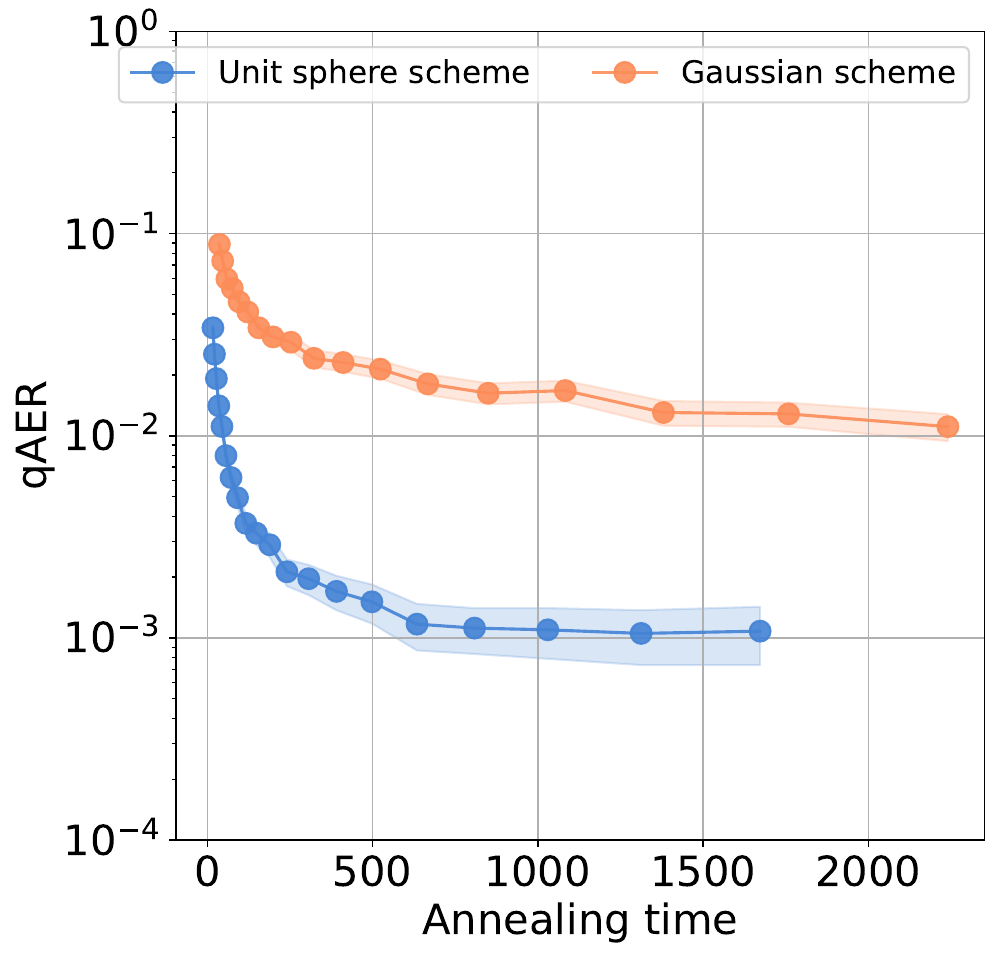}}
  \caption{The qAER reaches a plateau at $\text{SNR} = 5\text{dB}$ for both schemes (a,b) that is lowered when the precision level decreases. Hence we check at $\text{SNR} = 10\text{dB}$ the qAER against the annealing time (c) obtained by varying $\epsilon$} 
  \label{fig:qBER_evaluation}    
\end{figure*}

\subsection{Quantum activity-error rate}

The success probability is a vector-wise metric, however it would be more convenient to work with an element-wise one for signal processing applications. It is why we now introduce the \textit{quantum activity-error rate}.

Once the QA process reaches $\ket{\Psi(T)}$, one measures each quantum spin $\hat{\sigma}_i^z$ in this final state. We denote $\bm{\sigma}^{\text{QA}}$ the estimator of the activity pattern built by storing the result obtained for the $i$-th qubit in the associated component:
\begin{align}
    \sigma_i^{\text{QA}} = \left\{\text{outcome of the measurement of } \hat{\sigma}_i^z \text{ in } \ket{\Psi(T)} \right\}
\end{align}

However $\sigma_i^{\text{QA}} \in \{-1,1\}$, hence we rather start from the expression of the AER used in \cite{tanaka_statistical-mechanics_2002} which involves the correlation of the detector output $d$. A perfect recovery corresponds to $d = 1$ and the worst guess from the detector gives $d=-1$. Inspired by \cite{tanaka_statistical-mechanics_2002}, we introduce the \textit{quantum activity error-rate} of our detector:
\begin{align}
    \text{qAER} = \mathbb{E}_{(\bm{P},\bm{Y})} \left(\frac{1 - d}{2}\right),
    \label{qAER_def}
\end{align}
where $(1-d)/2$ corresponds to the proportion of the bit-error \cite{tanaka_statistical-mechanics_2002} for a given problem instance $(\bm{P},\bm{Y})$. In our case, $\bm{\sigma}^{\text{QA}}$ is a random variable due to the probabilistic nature of the quantum measurements. Its quantum expectation value is given by $\bra{\Psi(T)} \hat{\bm{\sigma}}^z \ket{\Psi(T)}$. Thus, we propose to define $d$ only through the quantum expectation value of the detector output:
\begin{align}
    d = \frac{1}{N} \sum_{i=1}^N \sigma_i^{(0)} \bra{\Psi(T)} \hat{\sigma}_i^z \ket{\Psi(T)}
\end{align}
We expect the quantum activity error rate to be lower-bounded by the activity error rate of the NNLS estimator. Indeed, our QA-based algorithm is supposed to converge towards $\bm{\sigma}^{\text{NNLS}}$. Thus, the estimator $\bm{\sigma}^{\text{QA}}$ is not supposed to beat the reliability of $\bm{\sigma}^{\text{NNLS}}$ on average. Since we fixed $\text{AER}^{\text{NNLS}}$, we expect that $\text{qAER} \geq 10^{-3}$.

\subsection{Reliability against the noise}

We first propose to check the behavior of the qAER metric against the SNR defined in Eq. \ref{snr_def}. To do so, we generated $N_{\text{samples}} = 10^4$ problem instances $(\bm{P},\bm{Y})$ for both distributions corresponding to the Gaussian pilot scheme and the Unit sphere scheme. 

Fig. \ref{subFig:qBER_vs_SNR_gaussian_N5M4} and \ref{subFig:qBER_vs_SNR_unit_sphere_N5M4} show the evolution of the qAER against the SNR with a $95\%$ confidence region for the estimation of the expectation value involved in Eq. \ref{qAER_def}. For both schemes, we simulate the behavior of a QA process with three precision levels $\epsilon \in \{0.1, 0.01, 0.001\}$ and compare it with $\text{AER}^{\text{NNLS}}$ as a function of the SNR. The Gaussian scheme appears to be more sensitive to the SNR than the unit sphere scheme. Furthermore, the reliability of the algorithm with the Gaussian scheme seems to be almost insensitive to the precision level at SNR $= 0$dB. Nevertheless, both curves rapidly reach a plateau located around $\text{SNR} = 10\text{dB}$ that corresponds to the lowest qAER one can reach at a given noise level. 

These results also show that the precision level required to reach some level of qAER strongly depends on the pilot scheme. Let us now check how it affects the annealing time required to ensure a given reliability of the algorithm.

\subsection{Annealing time}

Given a precision level $\epsilon$, we can evaluate the annealing time associated to the mean control function we use to parameterize a QA approach. For both pilot schemes, we generated some pairs $\left(\text{qAER}(\epsilon), T_{\text{mean}}(\epsilon)\right)$ with different values of the precision level. It allows to obtain the parametric curves of the qAER against the annealing time shown on Fig. \ref{subFig:qBER_vs_annealing_time_N5M4_snr10}.

Despite a larger confidence interval, the qAER of the unit sphere scheme reaches faster the lower bound of $10^{-3}$ than the Gaussian scheme. In terms of computational time, the Unit sphere scheme appears more interesting for our QA approach.

\section{Conclusion}

This paper tackles the AUD problem in massive wireless networks using a covariance-based approach. A non-negative least square strategy enables the implementation of a quantum annealing algorithm, known for its high reliability and reduced complexity in solving QUBO problems.

The choice of the matrix $\bm{P}$ appears to be an important decision in terms of computational time required by QA. The qAER metric converges faster towards $\text{AER}^{\text{NNLS}}$ for the unit sphere scheme than for the Gaussian scheme.

Having established the QA-based approach's sensitivity to pilot design, future work might focus on proposing preamble matrices $\bm{P}$ that meet stricter constraints. Using the compressed sensing framework can improve the likelihood of successful recovery for both the NNLS and QA estimators.

\section*{Acknowledgment}

We are grateful to Lélio Chetot, Maxime Guillaud for valulable discussions. This work was supported by the ANR under the France 2030 program, grant "NF-PERSEUS : ANR-22-PEFT-0004".

\nocite{*}
\AtNextBibliography{\footnotesize} 
\printbibliography

\end{document}